\documentclass[twocolumn,naturemag,superscriptaddress,lonbibliography]{revtex4-1}

\usepackage{natbib}
\usepackage{graphicx}
\usepackage{color}
\usepackage{amssymb}
\usepackage{amsfonts}
\usepackage{amsmath}
\usepackage{bm}
\usepackage{bbm}

\begin{document}

\title{Superconducting, Insulating, and Anomalous Metallic Regimes \\ in a Gated Two-Dimensional Semiconductor-Superconductor Array}
\author{C.~G.~L.~B\o{}ttcher}\thanks{Now at Harvard University, Cambridge, MA 02138, USA}
\affiliation{Center for Quantum Devices and Station Q Copenhagen, Niels Bohr Institute, University of Copenhagen, 2100 Copenhagen, Denmark}
\author{F.~Nichele}
\affiliation{Center for Quantum Devices and Station Q Copenhagen, Niels Bohr Institute, University of Copenhagen, 2100 Copenhagen, Denmark}
\author{M.~Kjaergaard}\thanks{Now at MIT, Cambridge, MA 02138, USA}
\affiliation{Center for Quantum Devices and Station Q Copenhagen, Niels Bohr Institute, University of Copenhagen, 2100 Copenhagen, Denmark}
\author{H.~J.~Suominen}
\affiliation{Center for Quantum Devices and Station Q Copenhagen, Niels Bohr Institute, University of Copenhagen, 2100 Copenhagen, Denmark}
\author{J.~Shabani}\thanks{Now at NYU, New York, NY 10003, USA}
\affiliation{California NanoSystems Institute, University of California, Santa Barbara, CA 93106, USA}
\author{C.~J.~Palmstr\o{}m}
\affiliation{California NanoSystems Institute, University of California, Santa Barbara, CA 93106, USA}
\affiliation{Department of Electrical Engineering, University of California, Santa Barbara, CA 93106, USA}
\affiliation{Materials Department, University of California, Santa Barbara, CA 93106, USA}
\author{C.~M.~Marcus}
\affiliation{Center for Quantum Devices and Station Q Copenhagen, Niels Bohr Institute, University of Copenhagen, 2100 Copenhagen, Denmark}

\date{\today}
\begin{abstract}
{\footnotesize The superconductor-insulator transition in two dimensions has been widely investigated as a paradigmatic quantum phase transition. The topic remains controversial, however, because many experiments exhibit a metallic regime with saturating low-temperature resistance, at odds with conventional theory. Here, we explore this transition in a novel, highly controllable system, a semiconductor heterostructure with epitaxial Al, patterned to form a regular array of superconducting islands connected by a gateable quantum well. Spanning nine orders of magnitude in resistance, the system exhibits regimes of superconducting, metallic, and insulating behavior, along with signatures of flux commensurability and vortex penetration. An in-plane magnetic field eliminates the metallic regime, restoring the direct superconductor-insulator transition, and improves scaling, while strongly altering the scaling exponent. }
\end{abstract}
\maketitle    
\indent

As temperature is lowered, two-dimensional (2D) materials may become superconductors, or, if highly disordered or in a strong magnetic field, may become insulators, with resistance diverging with falling temperature. Despite decades of work on this topic \cite{Jaeger1989, Lee1990, Goldman2010, Gantmakher2010, Vladimir2012}, the question of whether such a system can remain a metal, converging toward a finite resistance at low temperatures, remains unresolved \cite{KKS2017}. 

Within conventional frameworks, the 2D supercondutor-insulator transition (SIT) is driven either by phase fluctuations of the superconducting order parameter, leading to localized Cooper pairs, or reduction in the superconducting pairing amplitude due to unscreened repulsive interactions. In either case, theory suggests a direct quantum phase transition between superconductor and insulator as a function of disorder, magnetic field, or carrier density, with no metal phase. 

\begin{figure}

	\includegraphics[width= 2.5 in]{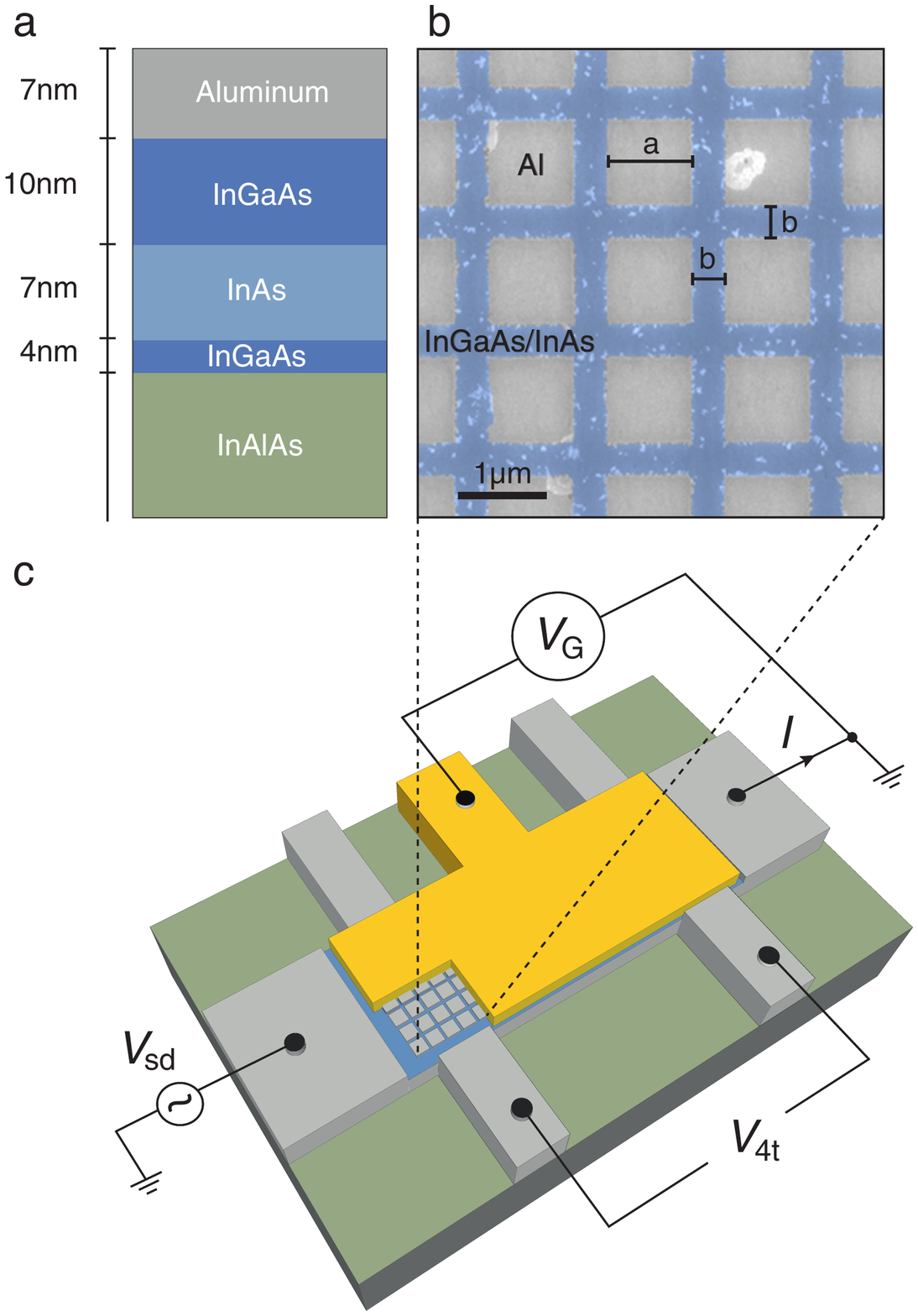}
	\caption{\textbf{Semiconductor/superconductor array}.  
			\textbf{a}, InGaAs/InAs heterostructure with epitaxial Al top layer. \textbf{b}, Scanning electron micrograph (false color) of device A before deposition of top gate, showing array of square Al islands with lateral dimension $a = 1\, \mu$m and separation $b = 150$~nm. Device B (not shown) has $b = 350$~nm. \textbf{c}, Device schematic showing currrent, $I$, source-drain voltage, $V_{\rm SD}$, and four-terminal voltage, $V_{\rm 4t}$, measured using side arms of Hall bar.
			 }
	\label{fig1}
\end{figure}

Much of the experimental literature has focussed on the direct SIT, often finding good agreement with scaling theory \cite{Gantmakher2010,Mason1999,Yazdani1995,Steiner2005,Bollinger2011,Schneider2012,Allain2012}. However, a number of experiments have reported an intervening metallic regime with saturating low-temperature resistance in a variety of systems, including metal films \cite{Markovic1999,Park2017}, oxides \cite{Steiner2008,Bollinger2011}, Josephson junction arrays (JJAs) \cite{Eley2012}, and superconducting islands on graphene, enabling electrostatic gating \cite{Han2014}. Proposed theoretical explanations for the observed metallic behavior include order-parameter fluctuations \cite{Feigelman1998,Spivak2001,SOK2008}, emergence of a Bose metal phase \cite{Lee1991, Kivelson1992, Phillips2003}, coupling to a dissipative bath \cite{Mason1999, Kapitulnik2001}, or a composite Fermi liquid phase \cite{Mulligan2016}.

Here, we experimentally investigate the transition from superconductor to insulator in a highly controllable system based on a two-dimensional electron gas (2DEG) formed in a semiconductor heterostructure with an epitaxial Al layer \cite{Shabani2016}. By depleting carriers between patterned islands of the Al layer with an electrostatic gate, the 2D sheet resistance, $R_{\rm s}$, was controlled from below 0.1~$\Omega$ to above $10^{8}~\mathrm{\Omega}$, spanning regimes of global superconductivity to insulating behavior, including a regime with anomalous low-resistance metallic behavior. This system, characterized by high mobility, large \textit{g}-factor, and strong spin-orbit coupling, along with large in-plane critical magnetic field of the Al islands, opens unexplored territories in condensed matter physics \cite{Chalker2001, Dimitrova2007}, including access to topological superconductivity and its transitions \cite{Levine2017}.

 \begin{figure}[t]

	\includegraphics[width= 3.3 in]{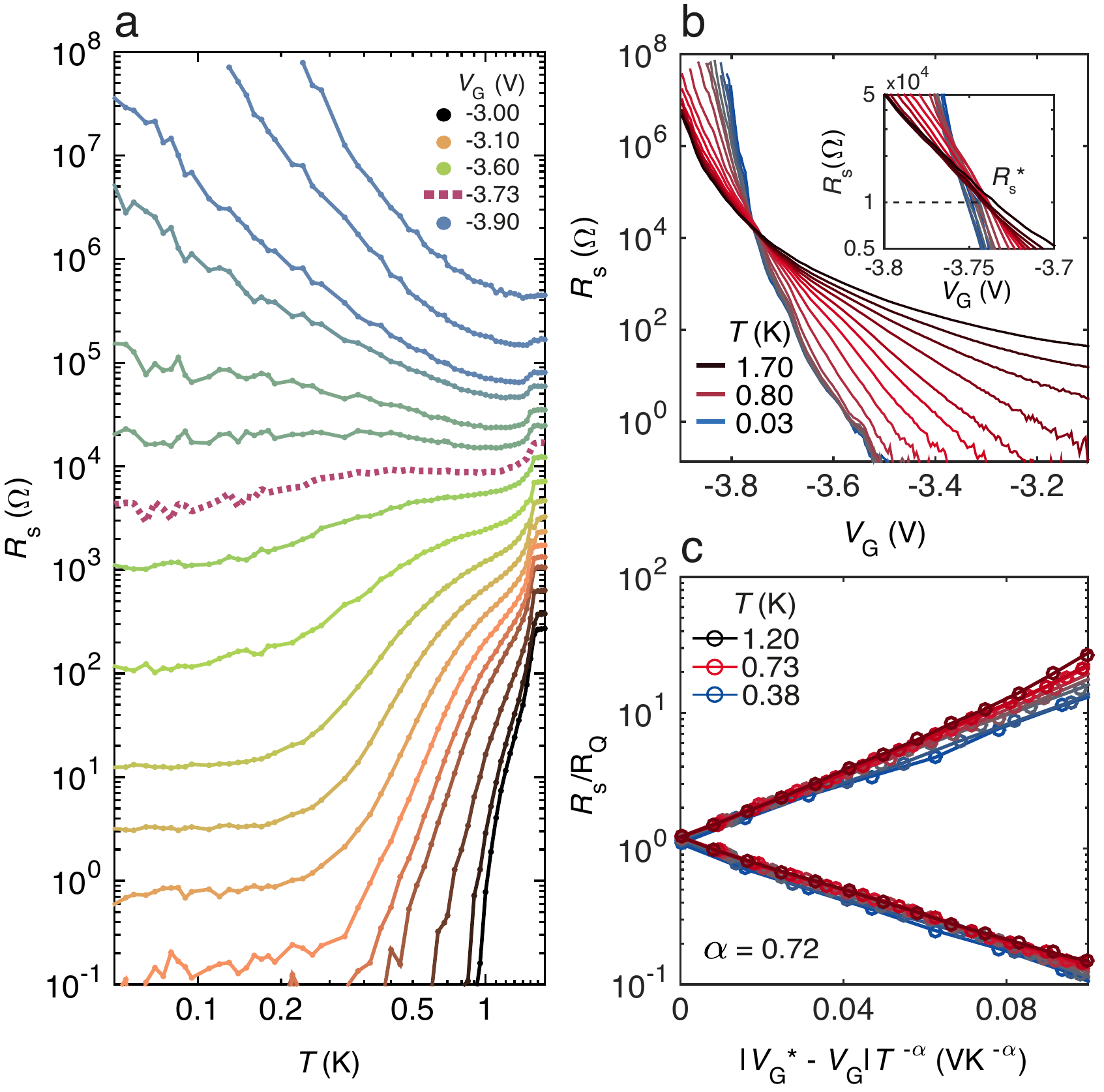}
	\caption{\textbf{Voltage controlled transitions}. 
			\textbf{a} Sheet resistance, $R_\mathrm{s}$, as a function of temperature, $T$, over a range of gate voltages, $V_{\mathrm{G}}$, from -3.0 V to -3.9 V, spanning the superconducting regime, anomalous metallic regime with saturating $R_\mathrm{s}$ at low temperatures, and insulating regime with $R_\mathrm{s}$ increasing with lower temperature. The kinks at $\sim 1.6$~K is where superconductivity vanishes in the Al. \textbf{b} $R_\mathrm{s}$ as a function of $V_{\mathrm{G}}$ for different temperatures reveals an approximate crossing point at $\sim$10 k$\Omega$. \textbf{c} Scaling plot yields best-fit exponent $\alpha = 0.72$, corresponding to  $z \nu = 1/\alpha = 1.4$. See text for details.}
	\label{fig2}
\end{figure}

The heterostructure material consist of a $7~\mathrm{nm}$ InAs quantum well, separated by a $10~\mathrm{nm}$ InGaAs barrier from a $7~\mathrm{nm}$ epitaxial Al layer, grown by molecular beam epitaxy \cite{Shabani2016}, as shown in Fig.~1(a). The Al layer was patterned into a 40$\times$100 array of square islands, each with lateral dimension $a = 1\,\mu$m. Two devices, A(B), with spacing $b = 150 (350)~\mathrm{nm}$ between islands [Fig.~1(b)], were measured. The spacing is comparable to the electron mean free path of $300~\mathrm{nm}$, extracted from measurements on a Hall bar with the Al removed. The structure was covered with $40~\mathrm{nm}$ of $\mathrm{Al_2O_3}$ insulator and a Ti/Au top gate [Fig.~1(c)]. 
Measurements were made in a dilution refrigerator with a base temperature of $25~\mathrm{mK}$ using standard lock-in techniques.  Both the current, $I$, through the device and the four-terminal voltage, $V_{\rm 4t}$, were directly measured using an AC voltage bias of $5~\mathrm{\mu V}$ or less [Fig.~1(c)] \cite{Chang2015,Kjaergaard2016}.

\pagebreak

\section{Gate-tuned transition and \\ anomalous metallic regime}

We use the dependence of $R_\mathrm{s}$ on temperature, $T$, at the lowest measured temperatures to distinguish insulating, metallic, and superconducting regimes. $R_\mathrm{s}(T)$ over a range of top-gate voltages, $V_{\mathrm{G}}$, is shown in  Fig.~2(a) for Device A. Device B showed similar behavior (see Supplementary Material). From positive to moderately negative gate voltages, $V_{\mathrm{G}}\gtrsim -3.1~\mathrm{V}$, $R_\mathrm{s}$ was unmeasurably small at the lowest temperatures, and at finite bias showed a critical current of a few $\mu$A (see Supplemental Information), indicating a global superconducting state. For more negative gate voltages, $-3.8\,V \lesssim V_{\mathrm{G}} \lesssim -3.1\,V$, $R_\mathrm{s}(T)$ saturated at the lowest measured temperatures to a gate-voltage-dependent value spanning three orders of magnitude in sheet resistance, from $R_\mathrm{s}\sim 1 \, \Omega$ to $R_\mathrm{s}\sim 1\,{\rm k}\Omega$, suggesting metallic behavior across this range. For even more negative gate voltages, $V_{\mathrm{G}} \lesssim  -3.8$ V, $R_\mathrm{s} (T)$ increased with decreasing temperature, indicating an insulating regime. At $V_{\mathrm{G}}^* \sim -3.7$ V, $R_\mathrm{s}$ was roughly independent of temperature, with $R_\mathrm{s}^* \sim 10\, {\rm k}\Omega$ in this region.  Within a conventional SIT framework, this can be identified as the separatrix dividing superconducting and insulating phases. The zero-field critical temperature $T_{\rm c}\sim 1.6$~K where superconductivity vanishes in the individual islands is visible as a kink at the high-temperature end of Fig.~2(a). As seen in Fig.~2(a)[S4(b)], for Device A[B], the lowest resistance curve that clearly saturates has a normal-state resistance of $\sim 2[6]$~k$\Omega$ above $T_{\rm c}\sim 1.6$~K.

Plotting $R_\mathrm{s} (V_{\mathrm{G}})$ for different temperatures yields a near-crossing of all curves around ${V_{\mathrm{G}}^*}\sim -3.75$ [Fig.~2(b)]. As seen in the inset of Fig.~2(b), the crossing is smeared over  a range spanning $\sim 10-20$ k$\Omega$. When restricted to temperatures above saturation, $T\gtrsim 0.3$~K, a much improved single crossing point is obtained.

Notwithstanding the metallic (saturating) behavior at low temperatures evident in Fig.~2(a), we perform a scaling analysis of these data within a conventional SIT framework. This will allow comparison to similar data at finite in-plane magnetic field, where the metallic phase vanishes, as discussed below. Scaling theory of the SIT \cite{Fisher1990} considers a correlation length $\xi$ and correlation time $\tau$,  both of which diverge at a critical tuning parameter, in this case gate voltage, with power-law dependences, $\xi \propto ({V_{\mathrm{G}}} -{V_{\mathrm{G}}^*})^{-\nu}$ and $\tau \propto \xi^z \propto ({V_{\mathrm{G}}} -{V_{\mathrm{G}}^*})^{-z\nu}$, yielding a scaling relation for sheet resistance,
\begin{align}
R_\mathrm{s}(V_{\rm G},T) = {R_\mathrm{s}^*}\, F(({V_{\mathrm{G}}}-{V_{\mathrm{G}}^*})T^{-1/z\nu}).
\end{align}
with a scaling function $F(x)$ that depends on distance from the critical gate voltage, $V_\mathrm{G}^{*}$, temperature, $T$, and the product of scaling exponents, $z\nu$.  

Experimentally, we plot $R_\mathrm{s}(V_{\rm G},T)$ as a function of $(V_{\rm G}-V_{\rm G}^*)T^{-\alpha}$,  treating $V_{\rm G}^{*}$ and $\alpha$ as fit parameters optimized to yield the best collapse of data across a range of temperatures and gate voltages. An alternative procedure \cite{Markovic1999, Bollinger2011} for extracting $\alpha$ from the temperature dependence of slopes $\partial R_{\rm s}/\partial V_{\rm G}$ at the critical point yielded consistent values (see Supplemental Material). Figure 2(c) shows $R_\mathrm{s}(V_{\rm G}, T)$ in units of $R_{\rm Q} \equiv h/4e^{2}$, demonstrating good scaling for $T\gtrsim 0.2$~K, covering four orders of magnitude of sheet resistance [Fig.~2(c)], with a best fit exponent $\alpha = 0.72$, corresponding to $z\nu = 1.4$.  Experimental scaling exponents are usually compared to theory based on classical percolation ($z\nu=4/3$) or quantum percolation ($z\nu\sim 2.3$)\cite{Steiner2008,Bollinger2011,Biscaras2013,Allain2012,Park2017}. The value we find is close to the classical percolation value, similar to what was found in a relatively clean disordered system \cite{Steiner2008}, where an intervening metallic regime was also observed. In that study, it was found that increasing disorder yielded a direct SIT with $R^{*}_{\rm s}\sim R_{\rm Q}$ and $z\nu\sim 2.3$, consistent with quantum percolation. 

\begin{figure}
	\includegraphics[width=3.5 in]{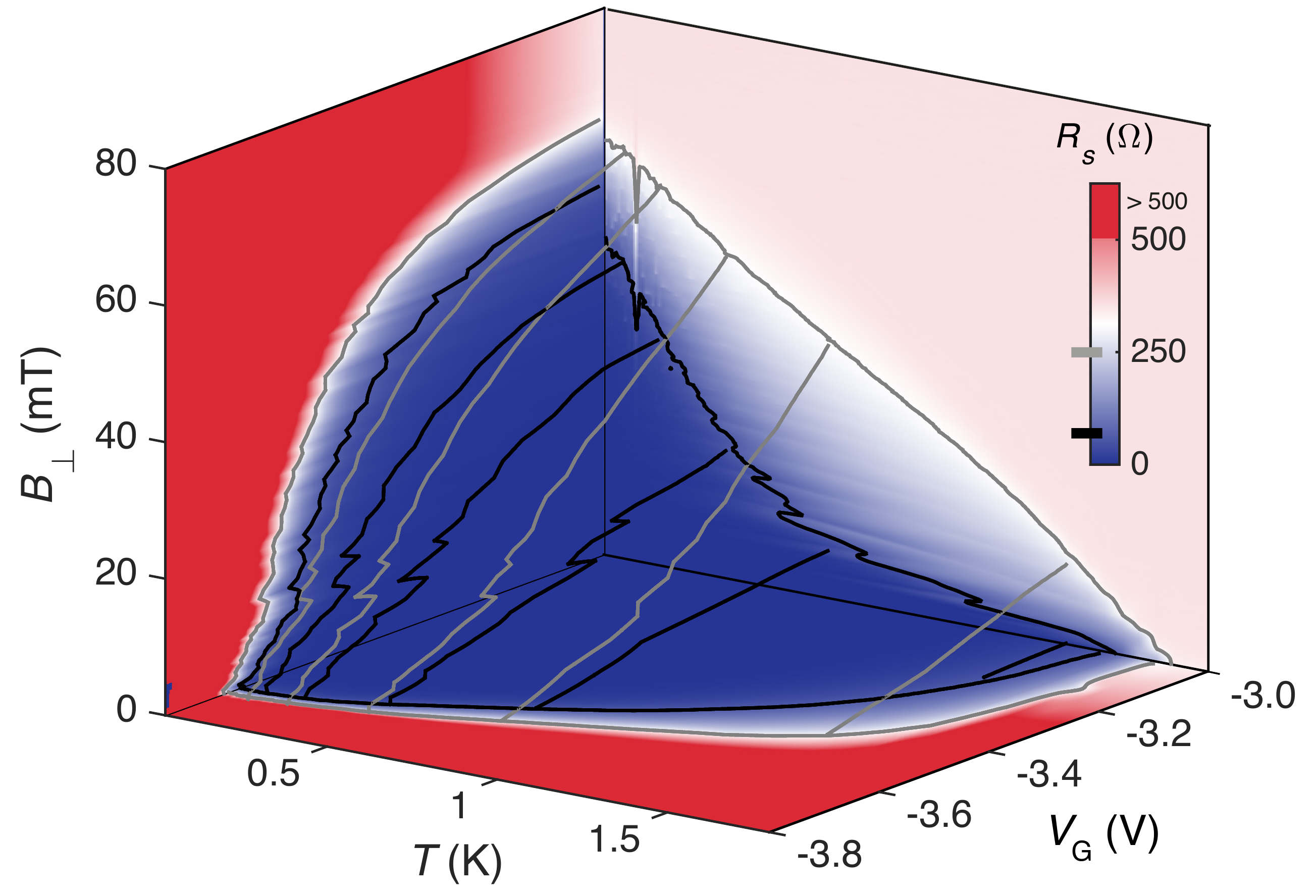}
	\caption{\textbf{Experimental phase diagram}. Three-dimensional phase-diagram-like plot of sheet resistance, $R_\mathrm{s}$ (color scale), as a function of temperature, $T$, perpendicular magnetic field, $B_{\perp}$, and gate voltage, $V_{\mathrm{G}}$. Contours at $50\:\Omega$ (black) roughly mark the boundary between superconducting (S) and anomalous metal (M$^{*}$) regimes. Contours at $250\:\Omega$ (gray) mark the critical critical field, $B_{c2}$, between anomalous (M$^{*}$) and normal metal (M) regimes in the $V_{\mathrm{G}}=-3.0\:\text{V}$ plane (back wall of diagram). Effects of a fourth control parameter, in-plane magnetic field, are shown in Fig.~6. }
	\label{fig3}
\end{figure}

\section{Phase diagram}

Motivated by the theoretical phase diagram for the dirty boson model of the SIT \cite{Fisher1990}, we show $R_{\rm s}$ as a function of $T$, $V_{\mathrm{G}}$, and perpendicular magnetic field, $B_{\perp}$, yielding the phase-diagram-like plot in Fig.~3. Rather than phase boundaries, the lines in Fig.~3 are contours of $R_{\rm s}$, with the $50~\mathrm{\Omega}$ contour roughly marking the boundary between superconducting (S) and anomalous metallic (M$^{*}$) regimes, and the $250~\mathrm{\Omega}$ contour marking the critical magnetic field $B^{\perp}_{c2}(V_{G}, T)$ to the normal metallic (M) regime for $V_{\mathrm{G}}=-3.0$~V, above which $R_{\rm s}$ is independent of temperature and field. The existence of two distinct crossovers in the $B_{\perp}$-$T$ plane is a robust experimental feature, investigated in detail below.

Plots of $R_{\rm s}$ in three $B_{\perp}$-$T$ planes are shown in Fig.~4, along with current-voltage ($I$-$V$) curves taken at indicated points in each of the planes. The plane in Fig.~4(a) at $V_{\rm G} = -3.0$~V is the same as the back wall of Fig.~3. A dashed curve in Fig.~4(a), repeated in Figs.~4(b) and 4(c), shows the boundary of M and M$^*$ regions. The boundary is clearly visible in the data in Fig.~4(a), less so in the other $B_{\perp}$-$T$ planes. At moderately negative gate voltages [Fig.~4(b)], the boundary to the superconducting regime is reentrant, with a dip around $T\sim 0.5\:\text{K}$. In the vicinity of this dip, two field-driven transitions are observed, an S-I transition on the low-temperature side and an S-M transition at higher temperatures. The boundary between these regimes, shown as a vertical dashed line in Fig.~4(b), is not visible in $R_{\rm s}$ but can be seen in the nonlinear I-V curves, described below.  At more negative gate voltages [(Fig.~4(c)] an insulating regime emerges at low temperatures, with resistance that first rises with $B_{\perp}$ then falls, similar to behavior in InO films \cite{Breznay2016}. 

 \begin{figure*}[ht]
 
 	\includegraphics[width=2\columnwidth]{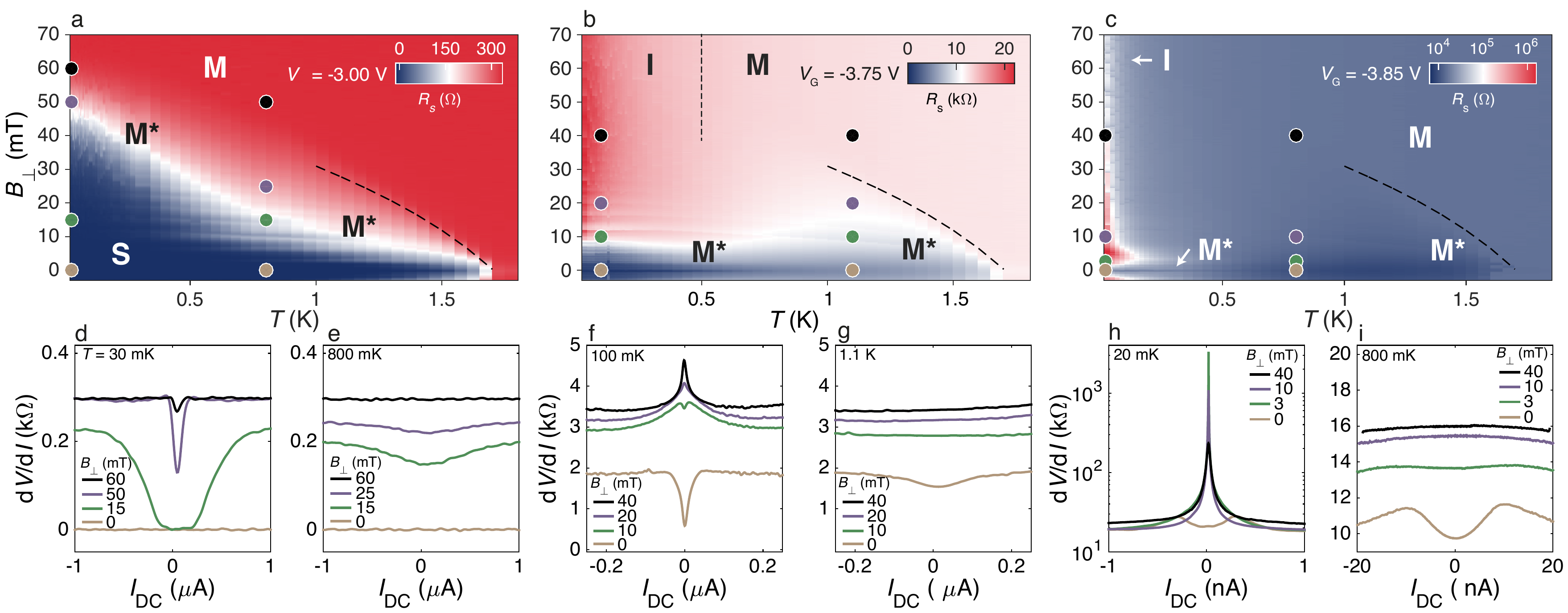}
 	\caption{\textbf{Temperature, magnetic field, and nonlinear response}. 
 			\textbf{a} Sheet resistance $R_{\rm s}$ (color scale) in the $T-B_{\perp}$ plane at the least negative gate voltage, $V_\mathrm{G}= -3.00\:\text{V}$ shows superconducting (S), anomalous (M$^{*}$) and normal (M) metal regimes. Boundaries between S and M$^*$ (white data) and between M$^*$ and M (dashed curve, determined by flat versus dipped $dV/dI$). The same dashed curve is also plotted \textbf{b} and \textbf{c}. \textbf{b} $R_{\rm s}$ (color scale) at intermediate gate voltage, $V_\mathrm{G}= -3.75\:\text{V}$. M$^{*}$ and I regimes determined from temperature dependence; M regime determined from the flat $dV/dI$ curves. Position of vertical dashed line separating I and M based on $dV/dI$ curves. Note the reentrant dip in the M$^{*}$ regime around 0.5 K. \textbf{c}  $R_{\rm s}$ (color scale) at more negative gate voltage, $V_\mathrm{G}= -3.85\:\text{V}$. M$^{*}$ shows a giant peak in $R_{\rm s}$ as a function of $B_{\perp}$, along with a narrow zero-bias peak in $dV/dI$ at 20 mK. The M$^{*}$ regime defined by temperature saturation, also associated with zero-bias dip in $dV/dI$. \textbf{d-i} Differential resistance $dV/dI$ as a function of applied current bias, $I_{\rm DC}$, at values of $T$, $B_{\perp}$, indicted by colored circles in \textbf{a-c}. }
 	\label{fig4}
 \end{figure*}

Further characteristics of the superconducing, insulating, anomalous metal, and normal metal regimes can be found in nonlinear transport.  Differential resistance $dV/dI$ as a function of applied dc current $I_{\text{DC}}$ is shown in Figs.~4(d-i), with the corresponding point where the measurement was taken indicated in the $B_{\perp}$-$T$ cuts, Figs. 4(a-c). Along the left axis of Fig.~4a ($T=30\:\text{mK}$), superconductivity was observed at small current bias, with a critical current of $I_c \sim 20\mu \mathrm{A}$ at zero field (Fig.~4d). Moving upward to $B_{\perp}=60\:\text{mT}$, superconductivity is lost, and normal metal (M) behavior observed, with $R_{\rm s}$ independent of $T$ and $B_{\perp}$, and flat $dV/dI$. At intermediate fields, the M$^*$ regime in Fig.~4(a), $dV/dI$ retains a dip around zero current, suggesting some vestige of superconductivity but not a zero resistance state. Moving to $T=800\:\text{mK}$ (Fig.~4e), the sampling of $dV/dI$ curves straddle the M$^*$--M boundary: In the M$^{*}$ regime at $B_{\perp} = 25\: \text{mT}$, $dV/dI$ showed a dip saturating at a nonzero value; In the M regime at $B_{\perp} = 40\:\text{mT}$, $dV/dI$ was flat.

Nonlinear transport at more negative gate voltages [Fig.~4(f)], crossing from M$^{*}$ to I, showed a zero-bias dip in $dV/dI$ at low field (M$^{*}$) and temperature, crossing to a zero-bias peak in $dV/dI$ at higher field (I). At the $B_{\perp}$-driven M$^{*}$-I crossover, ${R_\mathrm{s}}^*\sim 10\:\text{k}\Omega$ , consistent with the $T$-driven crossover described above [Fig.~2]. At temperatures above the reentrant feature around 0.5 K in Fig.~4(b), the peak in $dV/dI$ vanished, indicating a normal metallic (M) regime [Fig.~4(g)]. The vertical separation line in Fig.~4(b) is based on the vanishing peak in $dV/dI$, not by features in $R_{\rm s}$.

As the gate voltage is tuned farther negative, [Fig.~4(c)], the system exhibits strong insulating behavior at low temperature and magnetic field, with an associated giant zero-bias peak in $dV/dI$ as a function of field, spanning several orders of magnitude in resistance [Fig.~4(h)], as reported previously \cite{Steiner2005, Steiner2008, Breznay2016} and discussed in detail below. At elevated temperatures, $T > 800\: \text{mK}$, $R_{s}$ decreases, as expected in the insulating regime. At low fields, the zero-bias peak in $dV/dI$ seen at low temperature crosses over to a zero-bias dip, consistent with an M$^{*}$ regime, as  temperature is raised. Note that $R_{\rm s} > 10$~k$\Omega$, despite the zero-bias dip (Fig.~4i). 

Transport in the weakly insulating regime is well described by Efros-Shklovskii variable range hopping, $R(T)\propto {\rm exp}(T_1/T)^{1/2}$ \cite{Shklovskii1984}, as seen previously, for instance, in TiN films \cite{Baturina2007}. Fits yield $T_{1}\sim 2.5$~K, consistent with the theoretical prediction, $T_{\rm ES}= 2.8 (1/4\pi\epsilon' \epsilon_{0})(e^{2}/k_{\rm B} \xi)\sim 3\,K$, where $\xi$ is the coherence length and $\epsilon'$ is the material dielectric constant \cite{Shklovskii1984, Joung2012}. At the most negative gate voltages, transport crosses over to an activated form, $R(T)\propto {\rm exp}(T_0/T)$. Fits yield $T_{0}\sim 1.5$~K, comparable to the zero-field $T_{\rm c}$, a relation previously observed in disordered InO films \cite{Sambandamurthy2004}. Data for variable range hopping and activated transport are presented in the Supplementary Material. 
                                                        
\section{Flux effects}

We next investigate flux effects at low fields, both commensuration features associated with the periodic array, and flux penetration of individual Al islands. Both are evident in the three panels of Fig.~5, measured at different gate voltages. In Fig.~5(a), the features labeled $f=1...4$ indicate numbers of flux quanta per area of the periodic array, $f=B_{\perp}/B_0$, where $B_0 =\Phi_0/A$ with $A$ being the area of a unit cell [$A = (a+b)^2$] and $\Phi_0 = h/2e$ is the superconducting flux quantum. From the device design, $B_0 = 1.8\:\text{mT}$, consistent with the positions of these features. At integer values of $f$, there is a dip in $R_{s}$, which at the lowest temperatures extends into the superconducing state. Dips in $R_{\rm s}$ at fractional values of $f$, most prominently at $f=1/2$, are also observed.  Similar features are commonly observed in regular Josephson arrays \cite{Zant2001}. All panels in Fig.~5 are marked on their top axis with the positions of these features. They are most pronounced in Fig.~5(a), where superconductivity persists, and are essentially absent in the insulating regime [Fig. 5(c)]. Commensuration dips at integer and fractional $f$ values become less pronounced as the temperature is raised, but do not move in field with temperature.  In contrast, the positions of a second set of features in $R_{s}$, labeled $n=1...3$ in Fig.~5(a), are strongly temperature dependent. We associate this second set of features with vortex penetration of the individual Al islands, in good agreement with previous studies on single micron-size Al squares \cite{Baelus2006}.  

\begin{figure}
	\includegraphics[width=3.3 in]{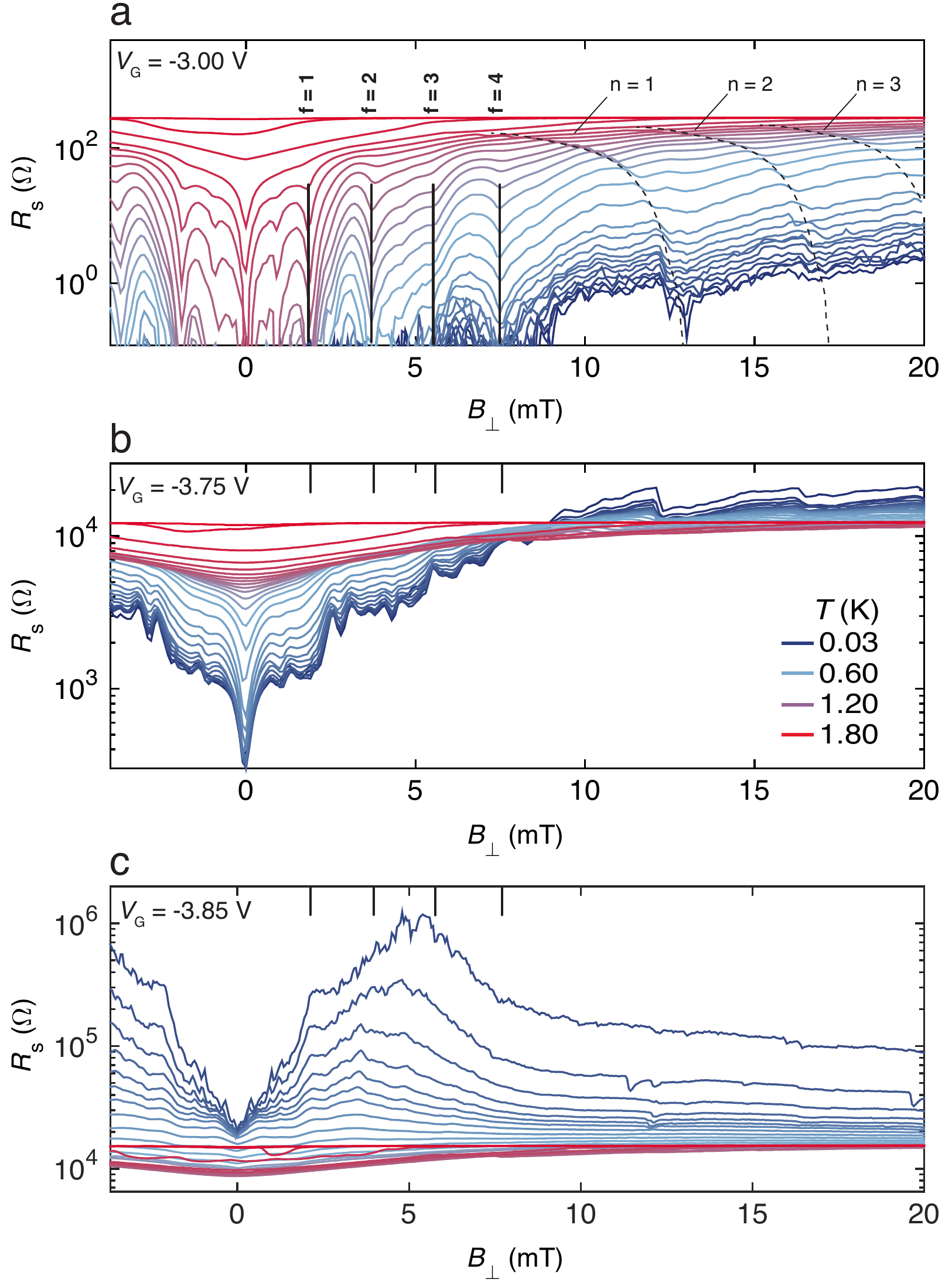}
	\caption{\textbf{Flux effects in S, M$^{*}$, and I regimes}. 
			\textbf{a} Sheet resistance, $R_{\rm s}$, in the superconducting regime, $V_\mathrm{G}= -3.00\:\text{V}$. Flux commensuration effects for $f=1...4$ flux quanta per unit cell of the array were independent of temperature (marked with vertical lines). Dips at fractional values $f=1/3, 1/2$ are also visible. Flux penetration into the Al islands for $n=1...3$ flux quanta per island occur at fields that depend strongly on temperature.  Legend for temperatures are given in \textbf{b}.  \textbf{b} At intermediate gate voltage, $V_\mathrm{G}= -3.75\:\text{V}$, low-temperature saturation of $R_{\rm s}$---the defining characteristic of the M$^{*}$ regime---is seen as an accumulation of curves at low-temperature. Saturation resistance shows mesoscopic fluctuations on a field scale smaller than one flux quantum per unit cell, with a large dip at zero field. At higher fields, the low-temperature curves exceed the high temperature curves above $\sim$10 mT, indicating a crossover to the insulating regime. \textbf{c} At more negative gate voltage, $V_\mathrm{G}= -3.85\:\text{V}$, in the insulating regime, a giant peak in $R_{\rm s}$ appears around a few times $B_0 =\Phi_0/A$, with some plateau structure at $B \sim B_{0}$ ($f \sim 1$). Legend for temperatures are given in \textbf{b}.}
	\label{fig5}
\end{figure}

Novel flux effects were observed in the M$^{*}$ regime, where $R_{\rm s}$ decreases with decreasing temperature then saturates [Fig.~5(b)]. Three features in the data are notable: (i) The low-temperature saturation value of $R_{\rm s}$ depends sensitively on magnetic field, showing mesoscopic fluctuations on a field scale smaller than $B_0$, with symmetric saturation values at positive and negative fields, and an absence of hysteresis that one might have anticipated for flux trapping.  (ii) The saturation value of $R_{\rm s}$ showed a deep minimum at $B_{\perp}=0$, the time-reversal symmetric situation. The field scale for the dip is less than 1 mT, smaller than $B_0$, suggesting that this feature is due to coherence among many islands. (iii) The saturation value of $R_{\rm s}$ ranges over two orders of magnitude, controlled by field, at fixed gate voltage. Lacking an understanding of the M$^{*}$ regime and the mechanism leading to saturation, we do not put forward a model of mesoscopic fluctuations and antilocalization of the saturation of $R_{\rm s}$.

The insulating regime [Fig.~5(c)] showed a giant peak in $R_{\rm s}$ at roughly the $B_0$ field scale, with a peak rising by nearly two orders of magnitude in a field of 5~mT at the lowest temperature. Similar effects have been reported in other SIT systems \cite{Sambandamurthy2004, Steiner2005, Steiner2008}. The field scale suggests that phase disorder in the array is responsible for the high resistance. A plateau-like feature at $\sim 2$ mT, seen over a range of temperatures, appears correlated with $f=1$ flux commensuration.

\section{In-plane magnetic field effects}

Because the epitaxial Al layer is thin (7 nm), the critical in-plane field is large, typically around 2~T \cite{Shabani2016}. However, due to both the finite thickness of the heterostructure and the large $g$-factor of InAs, global superconductivity in the arrays was found to vanish above a critical in-plane field $B_\parallel^*=0.6~\mathrm{T}$. We investigate effects of $B_\parallel$, oriented along the current direction, for Device B. An example for $B_\parallel=275~\mathrm{mT}$ is shown in Fig.~6, with additional data in the Supplemental Material. We found that the in-plane magnetic field largely suppresses the M$^{*}$ regime, leaving only a single separatrix moving horizontally throughout the temperature range, including the lowest temperatures. We speculate that the softening of the proximity-induced gap by the in-plane field \cite{Kjaergaard2016} quenched the coherence intrinsic to the anomalous metal regime, noting inconsistency with Ref.~\onlinecite{Kapitulnik2001}, which argued that dissipation stabilized the anomalous metal. Repeating the SIT scaling analysis, now in the absence of the intervening saturating curves, yields a single crossing point [Fig.~6(b)] of curves and better collapse of data [Fig.~6(c)]. We observed an unexplained systematic reduction of $\alpha$ with increasing $B_{\parallel}$, by over an order of magnitude, while the crossing point, ${R_\mathrm{s}}^*$, does not vary systematically [Fig.~6(b), inset]. Dependence of $\alpha$ on $B_\parallel$ is not understood, but may also result from dissipation due to the softening of the gap \cite{Wagenblast1997}.

\begin{figure}[!ht]
	\includegraphics[width=3.3 in]{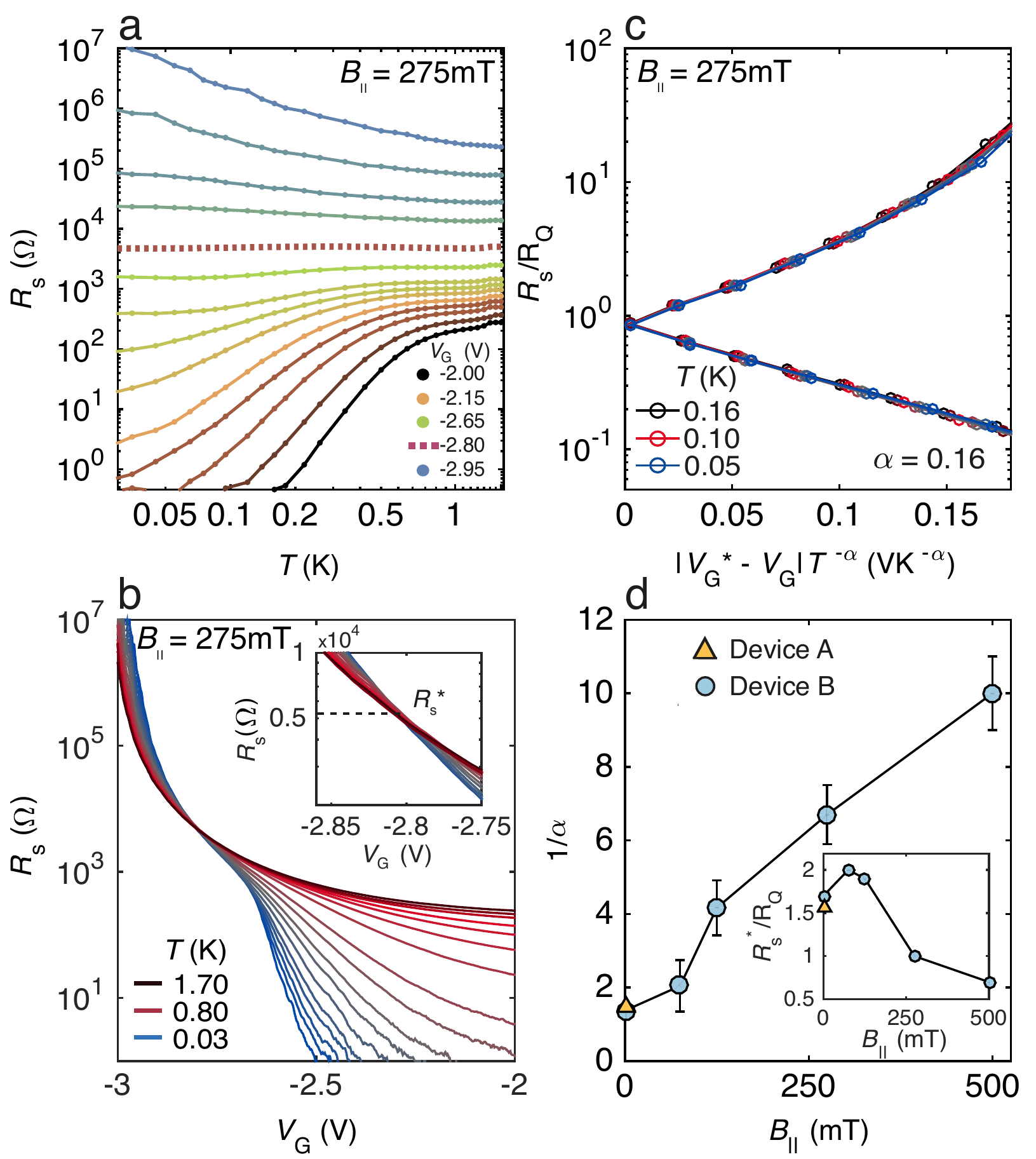}
	\caption{\textbf{In-plane magnetic field effects}.
			Gate-tuned SIT at $B_{\parallel}=275\:\text{mT}$. \textbf{a} Sheet resistance, $R_\mathrm{s}$, as a function of $T$ and gate voltage $V_{\mathrm{G}}$. Separatrix at $V_{\rm G}=-2.80\:\text{V}$ (dashed curve) corresponds to the intersection of curves in \textbf{b}.  \textbf{b} $R_\mathrm{s}$ as a function of gate voltage for several temperatures, shows very accurate single crossing point. Inset: Crossing point value of sheet resistance, ${R_\mathrm{s}}^*$, for several in-plane field values of device A (triangle) and B (circles) shows no clear trend. \textbf{c} Scaling plot (see text) for $B_{\parallel}=275\:\text{mT}$ shows excellent collapse with a single critical exponent $\alpha = 0.16$. \textbf{d} Critical exponent for device A (triangle, only at zero field) and B (circle) shows a clear trend of increasing $1/\alpha$ with $B_{\parallel}$. }
	\label{fig6}
\end{figure}

\section{Summary and Discussion}
The ability to repeatably and accurately control the crossover from superconducting to insulating behavior using a gate voltage enables new ways of  investigating the SIT and the long-debated anomalous metal state. A novel physical system comprising epitaxially grown semiconductor-superconductor heterostructures that makes these investigations relatively straightforward experimentally. In the present study, the device geometry itself is simple; we have not yet considered the wealth of new device geometries that become possible using range of materials and arbitrary lithographic patterning. The thinness of the epitaxial superconductor also makes in-plane field studies readily accessible, opening the door, for instance, to topological superconductivity \cite{Levine2017} and its transitions.  Several observations reported here in these initial experiments are familiar from the SIT \cite{Gantmakher2010, Vladimir2012} and JJ-array \cite{Zant1996} literature; some observations are new, accessible for the first time.

Besides the observed scaling in the vicinity of the critical $R_{\rm s}$, yielding a scaling exponent $z\nu \sim 1.4$, and a phase-diagram-like plot of $R_{\rm s}$ in the space of $T$, $B_{\perp}$, and $V_{\rm G}$, we observed an anomalous metallic regime, with saturating sheet resistance at low temperature, resembling similar behavior reported previously in a variety of systems \cite{Mason1999,Steiner2008,Allain2012,Couedo2016,Han2014,Eley2012}.  New observations concerning the metallic regime include mesoscopic fluctuations of the saturating resistance and a strong reduction in the saturation resistance at $B=0$, on a field scale significantly smaller than one flux quantum per period of the array. The absence of an anomalous metallic regime in an in-plane magnetic field indicates that it is unlikely that uncontrolled heating is responsible for the saturation. The striking dependence of the scaling exponent on in-plane field remains unexplained, but may result from dissipation \cite{Wagenblast1997} due to a softened gap \cite{Kjaergaard2016}.

\section{Methods}
The wafer structure was grown by molecular beam epitaxy to obtain $7~\mathrm{nm}$ InAs quantum well with a $10~\mathrm{nm}$ InGaAs upper barrier and $7~\mathrm{nm}$ of Al grown $in$ $situ$. The epitaxial grown Al results in a pristine superconductor/semiconductor interface  \cite{Shabani2016} and hard induced gap in the semiconductor \cite{Kjaergaard2016}.  

The devices were fabricated with first electron beam lithography to define the mesas with a Hall-bar geometry and etched using a wet etch solution of $\rm H_2\rm O:\text{citric acid}:\rm H_3 \rm P \rm O_4:\rm H_2 \rm O_2$ in the ratio $220:55:3:3$. The Al layer on each mesa was then selectively etched using Transene type-D to pattern it into a 40$\times$100 array of square islands \cite{Chang2015,Kjaergaard2016}. The results reported here are performed on Device A(B), with island dimension of $1~\mu\mathrm{m}$ and spacing between the islands was $150 (350)~\mathrm{nm}$. Atomic layer deposition of $40~\mathrm{nm}$ $\mathrm{Al_2O_3}$ was used to form a dielectric between the well-defined Al square array and a Ti/Au $(10/200~\mathrm{nm})$ top gate. Wire bonding to the epitaxial Al formed low resistance Ohmic contacts.

A separate Hall bar with all Al removed was used to characterize the 2DEG and partially depleted to enter a density regime with a single subband occupied, reached at a gate voltage value $V_{\rm G} \sim -2~\mathrm{V}$. Results yielding a maximum mobility of $\mu = 20,000~\mathrm{cm^2 V^{-1}s^{-1}}$ at a sheet density of $n_e = 1\times 10^{12}~\mathrm{cm^{-2}}$, corresponding to a mean free path $l_e=300~\mathrm{nm}$, comparable to the separation between islands. In addition, the epitaxial Al film has been characterized, yielding a superconducting transition temperature $T_c =1.5~\mathrm{K}$. In-plane and out-of-plane critical field values were measured to be $B_{||,c}\sim 1.6~\mathrm{T}$ and $B_{\perp,c}\sim 30~\mathrm{mT}$ respectively \cite{Suominen2017}.

Measurements were carried out in a dilution refrigerator with a base temperature of $25~\mathrm{mK}$. The four-terminal voltage, $V_{\rm 4t}$, and current, $I$, through the device were directly measured with AC voltage bias in all cases kept below $5~\mathrm{\mu V}$, using standard low frequency lock-in techniques. Series resistance in the lines were in the range of a few kilohms, creating an effective currrent bias in the low-resistance regime and voltage bias in the high-resistance regime.

In-plane field was precisely aligned with the plane of the array using a three-axis vector magnet by maximizing the critical current while trimming out-of-plane field coils.

\emph{Acknowledgements:} 
We thanks A.~Kapitulnik, S.~Kivelson, D.~Shahar, B.~Spivak, C.~Strunk, and V.~Vinokur for useful discussion. Research was supported by Microsoft Station Q and the Danish National Research Foundation. C.M.M. acknowledges support from the Villum Foundation. F.N. acknowledges support from a Marie Curie Fellowship (no. 659653).

\bibliography{Bibliography}

\setcounter{figure}{0}
\renewcommand{\thefigure}{S\arabic{figure}}
\newpage
\section{Supplementary Material}
\subsection{Scaling analysis}
The experimental scaling exponent, $\alpha$, used to collapse data in Figs.~2(c) and 5(c) of the main text, corresponds to $1/z\nu$ for the dirty-boson scaling theory of the superconductor-insulator transition (SIT) \cite{Fisher1990}, where $\nu$ describes scaling of correlation length and $z$ of correlation time, as discussed in the main text. The value of $\alpha$ was determined both by optimizing the collapse of the data and, following Ref.~\onlinecite{Shi2014}, by numerically differentiating $R_{\rm s}(V_{\rm G})$ in the vicinity of the critical gate voltage, $V_{\rm G}^{*}$, where curves at different temperatures cross at the critical sheet resistance $R_{\rm s}^{*}$ (see Figs.~2(b) and 5(b) in the main text).

\begin{figure}[b]
	\includegraphics[width = 7.5 cm]{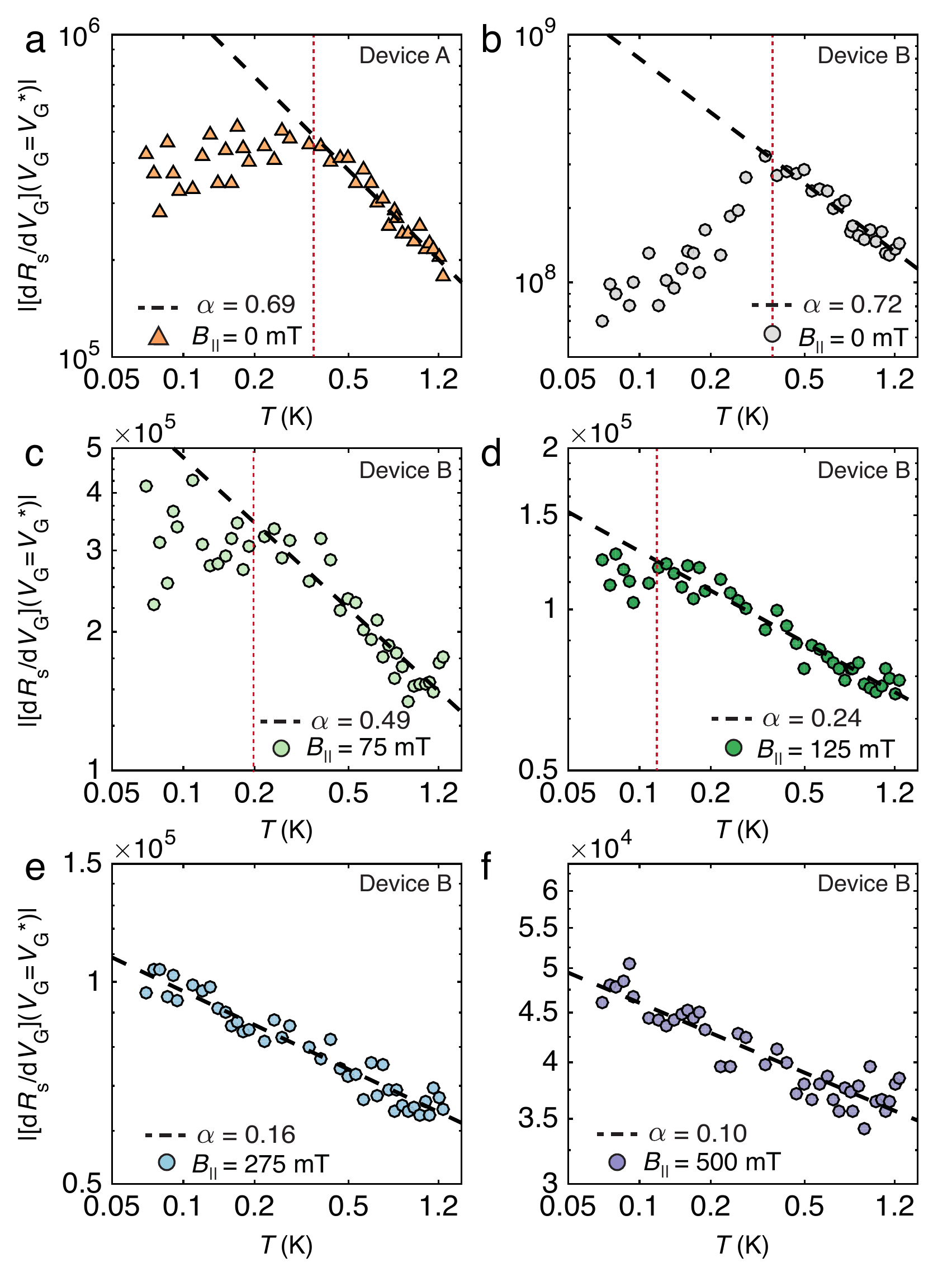}
	\caption{\footnotesize{\textbf{Scaling analysis}  Dependence on temperature, $T$, of the slope of sheet resistance, $R_{\rm s}$, as a function of gate voltage, $V_{\rm G}$, yields an estimate for the scaling exponent, $\alpha$. Note that scaling only works above a certain temperature (red dotted line), which decreases as an in-plane magnetic field, $B_{\parallel}$ is added, eventually covering the full temperature range of the measurement. The deviation from scaling presumably reflects the appearance of the anomalous metallic phase, M$^{*}$. Resulting values for $\alpha$ are plotted as a function of $B_{\parallel}$ in Fig.~6(d) of the main text.
	}}
	\label{figS1}
\end{figure}

Fitting a line to the log of the absolute value of the derivative at the critical point, $|\partial{R_{\rm s}(T,V_{\rm G}})/\partial{V_{\rm G}}|_{V_{\rm G}=V_{\rm G}^{*}}$, as a function of temperature, $T$, yields a value for $\alpha$ as the slope of that best-fit line,
\begin{align}
	\log |\partial{R_{\rm s}(T,V_{\rm G}})/\partial{V_{\rm G}}|_{V_{\rm G}=V_{\rm G}^{*}} = \log|{R_{\rm s}^*F'(0)|} -\alpha \log(T).
\end{align}

Data and best-fit lines are shown in Fig. S1 for several values of in-plane magnetic field ranging from $B_{\parallel} =0$ to $B_{\parallel} = 0.5$ T. For $B_{\parallel} =0$ in both devices, data at higher temperatures are well described by the line fit, while at lower temperatures, the data deviates significantly, associated with the onset of the anomalous metallic regime (M$^{*}$). In these cases, the value of $\alpha$ was determined by fitting the data above crossover temperature, out of the M$^{*}$ regime, where the fit is good, marked as a vertical dotted line in Fig.~S1. Figures~S1(c-f) demonstrate how the application of $B_{\parallel}$ shrinks the M$^{*}$ regime, extending the temperature range where scaling works well to the lowest measured temperatures.

\subsection{Comparing superconducting, anomalous metallic, and insulating regimes for Devices A and B}
The three $T$-$B_{\perp}$ planes in Fig.~4 of the main text show S, M$^{*}$, and I regimes at low temperature and field, and a normal metallic (M) behavior at high temperature and field. These plots, for Device A, are shown again in Fig.~S2 without marked boundaries or labels, along with corresponding data for Device B. Gate voltages for Device B were set so that its normal-state resistance matched the normal-state resistance of Device A in each of the planes. Behavior of Device A and Device B are similar, despite the difference in island space in the two devices ($b_{\rm A} =$ 150 nm, $b_{\rm B} =$ 350 nm). 

Dashed curves in Fig.~4 (main text) indicating the crossover from M and M$^*$, can now be more readily seen a contrast in the data in Fig.~S2(a), which is the same data as Fig.~4(a), with no curve added. The curve Fig.~4(a) is a fourth order polynomial fit to the contrast feature. The corresponding curves are less visible in Figs.~S2(b,c). The gate voltages for Device A are the same as shown in Fig.~4. 

\begin{figure*}
	\includegraphics[scale = 0.4]{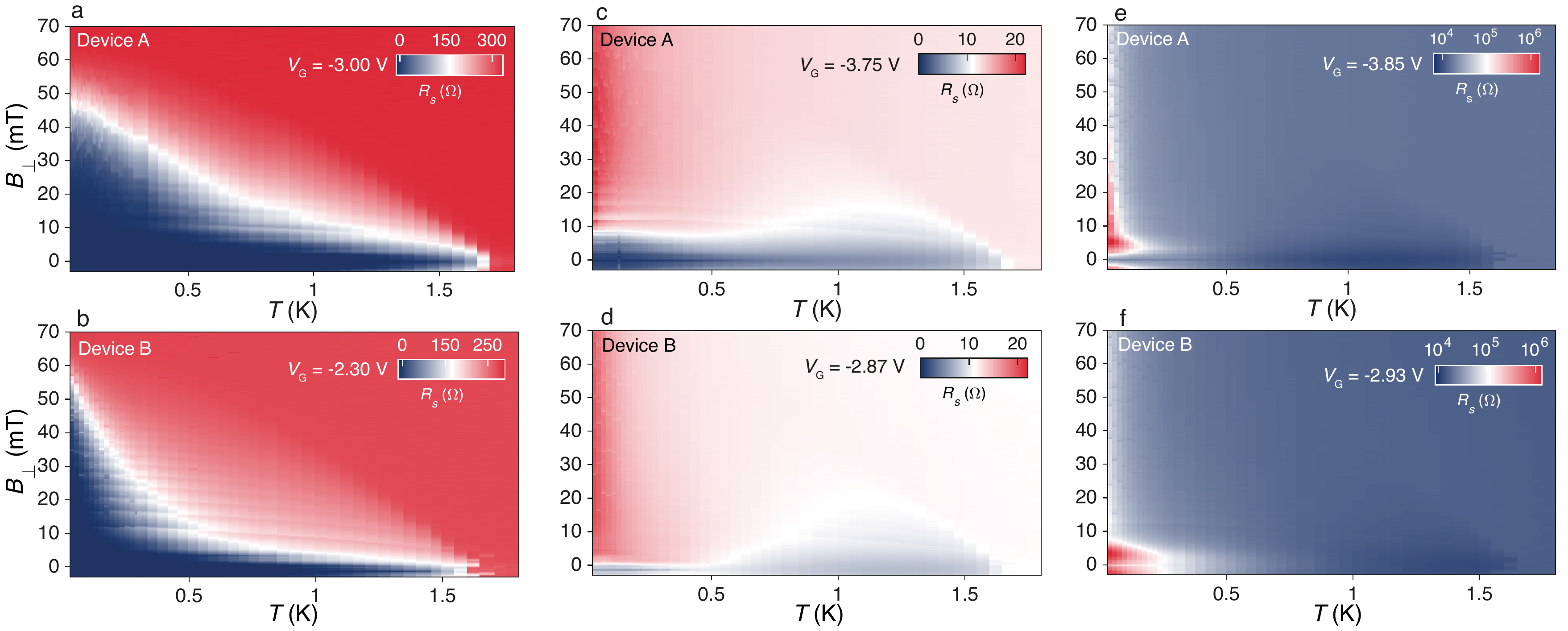}
	\caption{\footnotesize{\textbf{Comparing temperature-field planes for two devices} {\bf a-b} Two-dimensional plots of sheet resistance, $R_{\rm s}$, as a function of temperature, $T$, and perpendicular magnetic field, $B_{\perp}$  for Devices A and B, measured at gate voltages that gave equal normal state resistances in the two devices, spanning the superconducting (S) and anomalous metal (M$^{*}$) regimes, following labels in Fig.~4(a) of the main text. {\bf c-d} Temperature-field planes for Devices A and B at comparable, moderately negative, gate voltages showing insulating (I), anomalous metal (M$^{*}$), and normal metal (M) regimes, following the labels in Fig.~4(b) in the main text. {\bf e-f} Temperature-field planes for Devices A and B at comparable, more negative gate voltages, showing insulating (I) regime at  low temperature, weakly anomalous metal (M$^{*}$), and normal metal (M) at higher temperature, following the labels in Fig.~4(c) in the main text. 
	}}
	\label{figS2}
\end{figure*}

\begin{figure}
	\includegraphics[width = 7.5 cm]{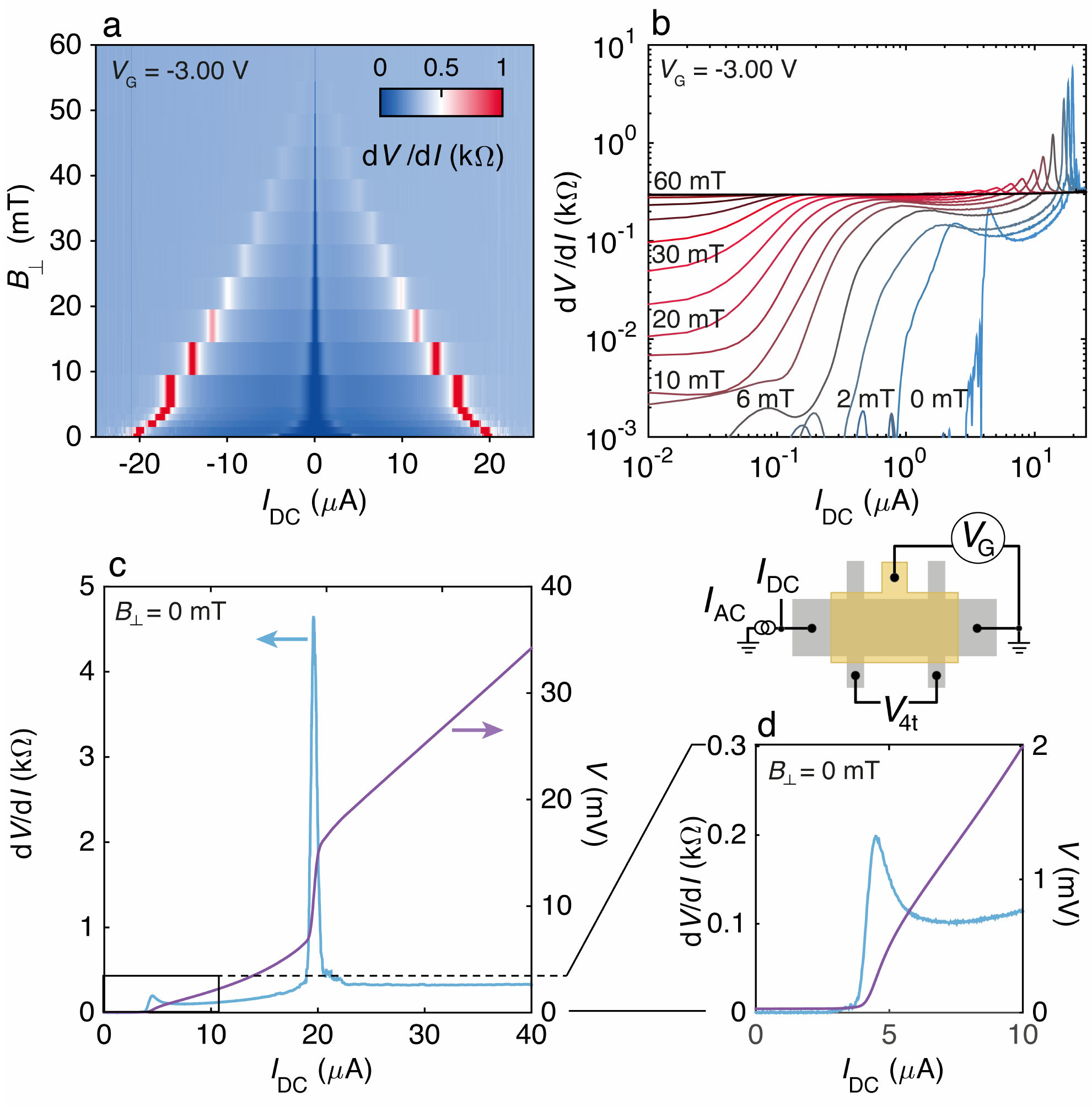}
	\caption{\footnotesize{\textbf{Critical currents in superconducting regime}. {\bf a} Differential sheet resistance, $dV/dI$, as a function of dc current bias, $I_{\rm DC}$ and perpendicular magnetic field, $B_{\perp}$, showing two critical currents at $B_{\perp}=0$ from global superconductivity to flux-flow to normal. Discrete steps in $B_{\perp}$ result from coarse spacing of field values. {\bf b} Same data as {\bf a} with log-log scale shows nonzero $R_{\rm s}$ above $B_{\perp} \sim 6$~mT. {\bf c} Differential sheet resistance (left axis) and voltage drop per square (right axis) at $B_{\perp}=0$, showing two critical currents. {\bf c} Zoom-in on first transition from zero-resistance to low-resistance (flux flow) state. A schematic of the measurement set up is also shown.
	}}
	\label{figS3}
\end{figure}

\subsection{Current bias dependence: critical currents separating zero-resistance, flux-flow, and normal-state transport}
Figure~4(d) of the main text shows a zero-resistance state at zero magnetic field in the range of $\pm 1\,\mu$A current bias. On an expanded range, two transitions to resistive states were observed, defining two critical currents, as shown in Fig.~S4 for Device A for $V_{\rm G} =-3.00$~V. Below the first transition, $dV/dI$ was unmeasurably small, suggesting global superconductivity, corresponding to the zero-field data in Fig.~4(d) of the main text.

As seen in Fig.~S3(b), a first transition occurred at $\sim 5\,\mu$A into a finite-resistance state. In this state, the voltage  increased roughly in proportion to the applied current bias. That is, the normalized differential resistance $dV/dI$ was roughly constant. Normalization of $dV/dI$ is done by multiplying by $W/L = 0.32$, where $W$ is the width of the Hall bar and $L$ is the distance between voltage probes. With this normalization, $dV/dI$ is a differential {\it sheet} resistance, corresponding to $R_{s}$ for the case of zero bias. As seen in Fig.~S3(b), at $\sim 20\,\mu$A a second transition was observed to a state with larger differential sheet resistance, $dV/dI$.  The value of $dV/dI$ above the second transition is equal to $R_{s}$ in the normal metallic (M) regime. We associate the lower-$dV/dI$ state with the flux-flow regime \cite{Orlando1991}, noting the increasing $dV/dI$ with increasing $B_{\perp}$, characteristic of flux-flow resistance. Both critical currents depend on magnetic field, the higher one decreasing roughly linearly with $B_{\perp}$, as seen in Fig.~S3(a).

\subsection{Full data sets with in-plane magnetic field}

Application of an in-plane magnetic field, $B_{\parallel}$, reduces the crossover temperature to the anomalous metallic (M$^{*}$) regime, where scaling fails [see Fig.~S1] and sheet resistance $R_{\rm s}(T)$ saturates. In-plane field improves the overall quality of scaling and lowers the scaling exponent, $\alpha$. In the main text, these trends are presented in Fig.~6. Figure~S4 shows full data sets of $R_{\rm s}(T, V_{\rm G})$ at different values of $B_{\parallel}$. Devices A and B are shown at $B_{\parallel}=0$. For other values of $B_{\parallel}$, Device B is shown.

\begin{figure}
	\includegraphics[width = 7.5 cm]{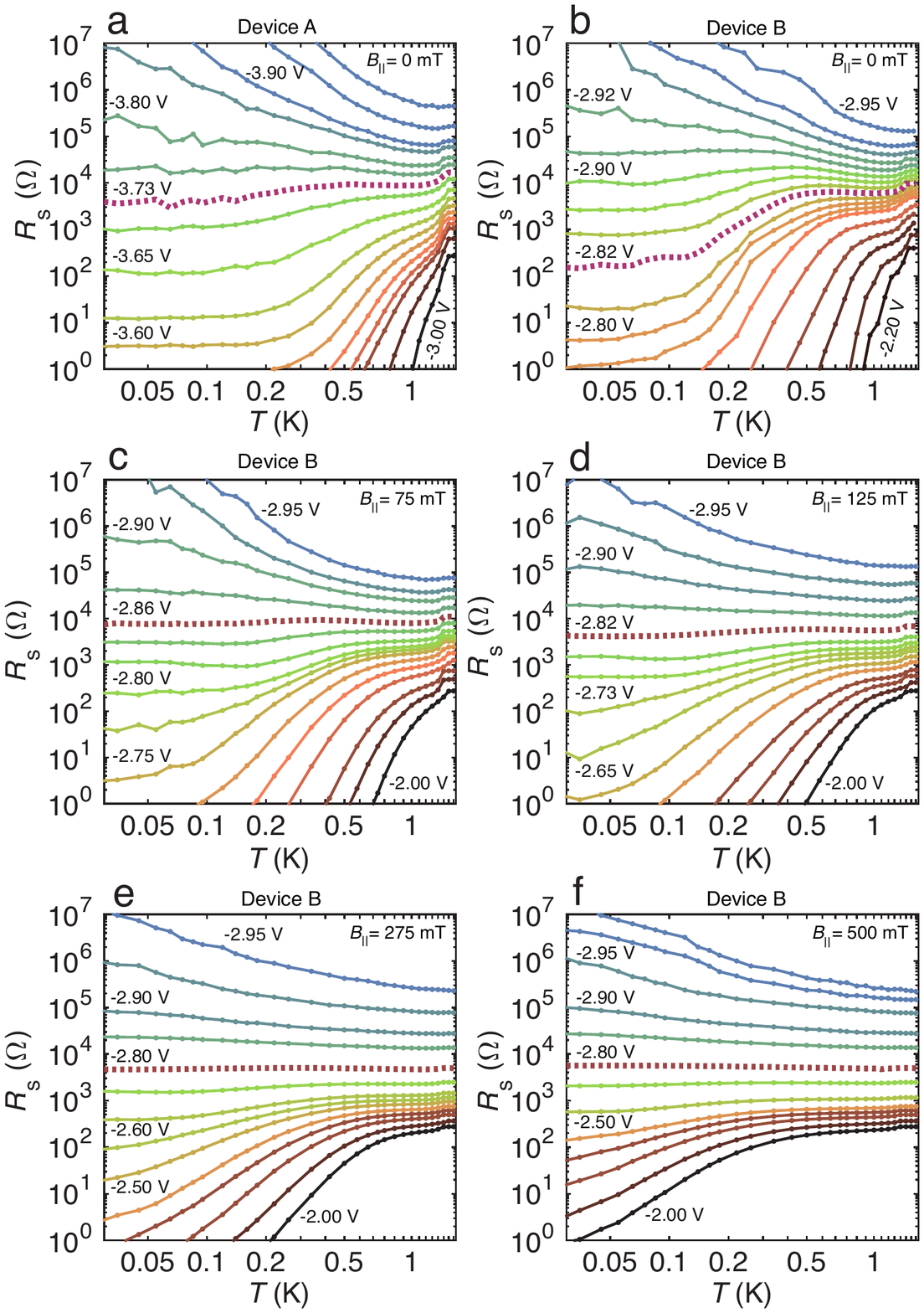}
	\caption{\footnotesize{\textbf{In-plane magnetic field effects on SIT transition} {\bf a} Temperature evolution of sheet resistance, $R_{\rm s}(T)$ at $B_{\parallel}=0$ for Device A with gate voltages indicated, showing S, M$^{*}$, and I regimes and separatrix (dashed curve). Points along the curve show measured values. {\bf b} Same as  {\bf a}, except for Device B. Note difference in gate voltage range. Separatrix based on flat curve at higher temperature, $T\gtrsim 0.3$ K. {\bf c-e} Sheet resistances, $R_{\rm s}(T)$, for Device B at in-plane fields $B_{\parallel}=$ 75 mT, 125 mT, 275 mT, and 500 mT shows lowering of the saturation temperature.
	}}
	\label{figS4}
\end{figure}

\subsection{Insulating regime: activated transport and variable range hopping}
We characterize transport on the insulating side of the gate-tuned SIT [Fig.~S5(a)] by constructing two plots that each displays the temperature dependence of the resistance. Deep in the insulating regime, roughly from $V_{\rm G} = -3.85~\mathrm{V}$ to $-3.90~\mathrm{V}$ the resistance show activated behavior, $R(T)\propto {\rm exp}(T_0/T)$, and follows a straight line on Arrhenius plot [Fig.~S5(b)]. At lower resistances but still in the insulating regime, transport does not look activated; the three lower curves in Fig.~S5(b) appear curved. Instead, in the weakly insulating regime near the separatrix, transport appears well described by Efros-Shklovskii variable range hopping (ES-VRH), $R(T)\propto {\rm exp}(T_1/T)^{1/2}$ \cite{Shklovskii1984}. This can be seen in Fig.~S5(c), where the lower three curves appear as straight lines when $R_{\rm s}$ plotted on a log vertical scale against $T^{-1/2}$.  The existence of a smooth crossover from ES-VRH to activation suggests that the insulating regime is first accompanied by a suppression of the density of states around zero energy due to Coulomb interactions, which at more negative gate voltages becomes a Coulomb gap. For ES-VRH, fits in Fig.~S5(c) give $T_{1}\sim 2.3-2.8$~K, depending slightly on gate voltage, in good agreement with the theoretical estimate \cite{Shklovskii1984, Joung2012},

\begin{align}
	T_{\rm ES}= 2.8 (\frac{1}{4 \pi\epsilon' \epsilon_{0}})(\frac{e^{2}}{k_{\rm B} \xi})\sim 3\,{\rm K}.
\end{align}

The estimate 3~K is obtained by taking the coherence length, $\xi$, as the size of the plaquette, $\xi \sim a = 1 \mu$m, and the effective dielectric constant $\epsilon' \sim 15$ as the average of InAs ($\epsilon' = 11$) and HfO$_{2}$ ($\epsilon' = 18$). 

In the activated regime, fits in Fig.~S5(b) give $T_{0}\sim 1.5$~K. This value is close to the zero-field critical temperature, $T_{\rm c}\sim1.6$~K. The similarity of values for $T_{0}$ and $T_{\rm c}$ was noted previously in amorphous InO films \cite{Sambandamurthy2004}. Previous studies on disordered TiN films \cite{Baturina2007} showed  a crossover from activated transport to ES-VRH as the temperature was increased, not as a function of disorder. Values for $T_{1}$ in \cite{Baturina2007} were similar to the values found here, while values for $T_{0}$ were three to five times smaller than found here.

\begin{figure}[t]
	\includegraphics[width = 7 cm]{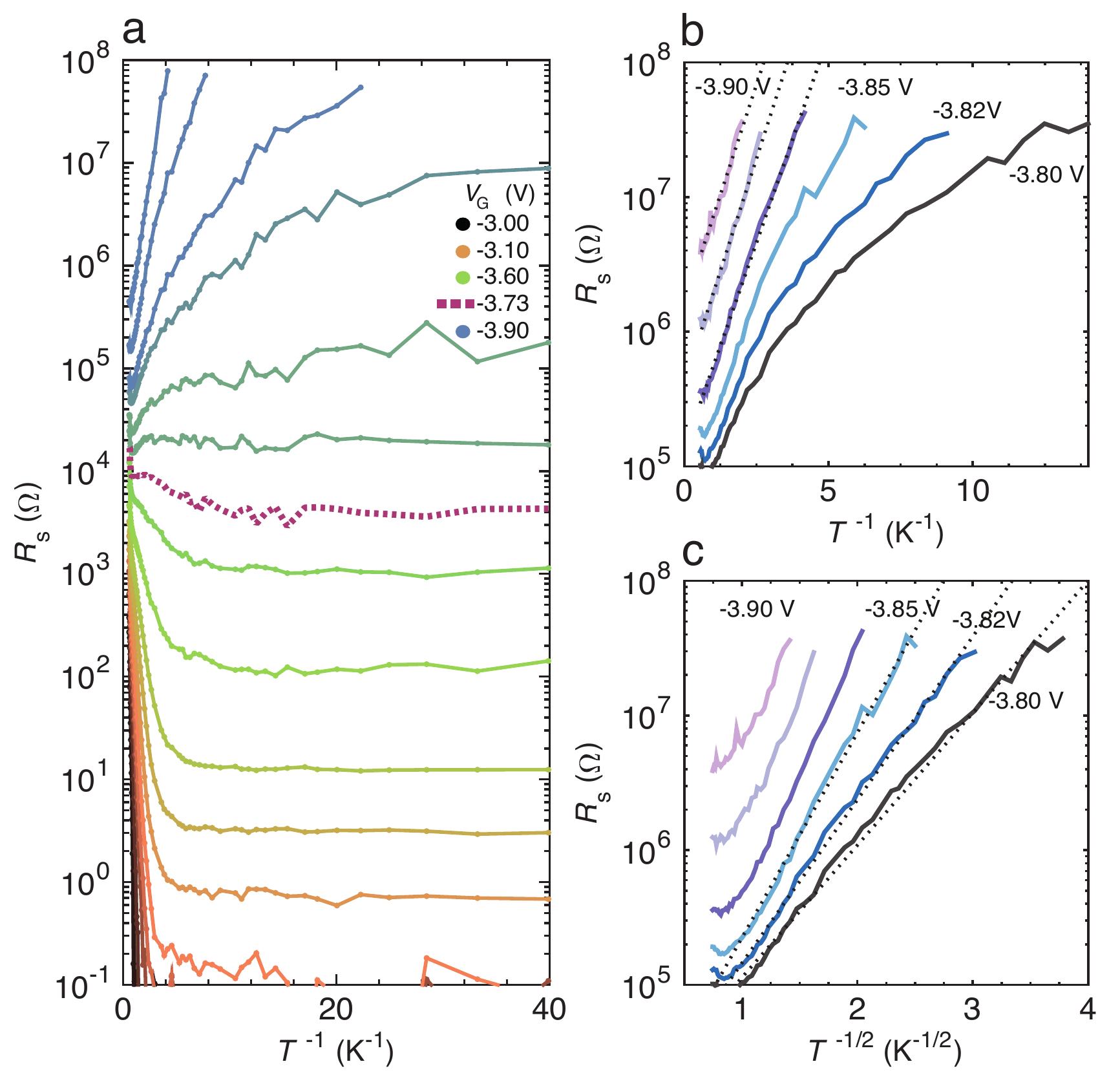}
	\caption{\footnotesize{\textbf{Activated transport and variable-range hopping}. {\bf a} Same data as Fig.~2(a) versus inverse temperature with log-log axes. {\bf b}, Same data as {\bf a}, with a linear horizontal axis. A straight line on this plot indicates activated transport, $R(T)\propto {\rm exp}(T_0/T)$. Line fit to most negative gate voltages (dashed lines) yield $T_{0}$~=~1.5 K.  {\bf c} Same data as {\bf b}, plotted versus $T^{-1/2}$. Efros-Shklovskii variable range hopping (ES-VRH), $R(T)\propto {\rm exp}(T_1/T)^{1/2}$ appears as a straight line. The three bottom data sets (dashed lines) yield $T_{1}$~=~2.3~K, 2.5~K, 2.8~K for $V_{\rm G}$ = $-$3.80~V, $-$3.82~V, $-$3.85~V. 
	}}
	\label{figS5}
\end{figure}

\end{document}